\documentclass[reprint,superscriptaddress,showkeys,showpacs]{revtex4-2}
\usepackage{verbatim}
\usepackage{color}
\usepackage{subfigure}
\usepackage[]{graphicx}
\usepackage{dcolumn}
\usepackage{bm}
\usepackage{blindtext}
\usepackage{floatrow}
\usepackage{makeidx}
\usepackage{epsfig}
\usepackage{hyperref}
\usepackage{array}
\usepackage{epstopdf}
\usepackage{braket}
\usepackage{multirow}
\usepackage{mathtools}
\usepackage{amsmath,amssymb,amsfonts}
\usepackage{fancyhdr}
\usepackage{multirow}
\newcommand{\RNum}[1]
{\uppercase\expandafter{\romannumeral #1\relax}}
\newcommand{\Rnum}[1]
{\lowercase\expandafter{\romannumeral #1\relax}}
\def\qe{\textsc{Quantum ESPRESSO}}

\makeindex
\begin{document}
\preprint{APS/123-QED}
\title{Electronic structure of SLSiN under charge density modulation}
\author{Ashkan Shekaari}
\email{shekaari@email.kntu.ac.ir}
\email{shekaari.theory@gmail.com}
\affiliation{Department of Physics, K. N. Toosi University of Technology, Tehran 15875-4416, Iran}
\date{\today}
\begin{abstract}
First-principles density-functional theory calculations were carried out to assess how incremental unit-cell charging alters the electronic behavior of SLSiN (single-layer Si$_3$N$_4$). The net charge per cell was systematically tuned from $n\,=\,0$ (the neutral/reference configuration) to $n\,=\,\pm\,1,\pm\,2$, and $\pm\,3$ elementary charges, and for each charged configuration the band structure and density of states were evaluated at the PBE level. In its neutral state, SLSiN exhibits zero electronegativity, signifying both its indifference to additional electron density and its intrinsic stability when integrated into heterostructures. Altogether, these results reveal that precise control of the charge density can drive SLSiN across an insulator-to-metal transition.
\end{abstract}
\maketitle
\subsection*{{\normalsize{I. Introduction}}}
Silicon nitride (Si$_3$N$_4$) is a synthetic ceramic first prepared in 1859 and initially molded into refractory components during the 1950s and 1960s~\cite{1}. Its development as a high-performance engineered ceramic accelerated throughout the 1980s~\cite{2,3,4,5}. Owing to its primarily covalent bonding, Si$_3$N$_4$ uniquely combines outstanding mechanical strength and toughness at both ambient and elevated temperatures with excellent oxidation and thermal-shock resistance~\cite{6,7,8,9,10,11,12,13}. These exceptional properties have driven extensive research to deliver ceramics capable of enduring heavy loads, severe wear, and corrosive environments~\cite{14}.

Among the silicon nitride phases including Si$_3$N$_2$, Si$_2$N$_3$, and Si$_3$N$_4$~\cite{15}, the latter (Si$_3$N$_4$) is the most thermodynamically stable polymorph. This exceptional stability underlies its widespread deployment in high-performance applications, including turbine components~\cite{16}, ball and roller bearing elements~\cite{17,18}, metal-working and cutting tools~\cite{19}, thermocouple tubes and molten-metal crucibles~\cite{20}, semiconductor fabrication~\cite{21,22}, AFM cantilevers~\cite{23}, and mechanical seals~\cite{24,25}. 

In addition to such industrial applications, Si$_3$N$_4$'s biocompatibility~\cite{bioc} and mechanical resilience have spurred its adoption in orthopedic implants. It is now under consideration as an alternative to PEEK (polyether ether ketone) and titanium in spinal fusion devices~\cite{26,27,28,29}.

\begin{figure}[H]
	\centering
	\includegraphics[scale=0.47]{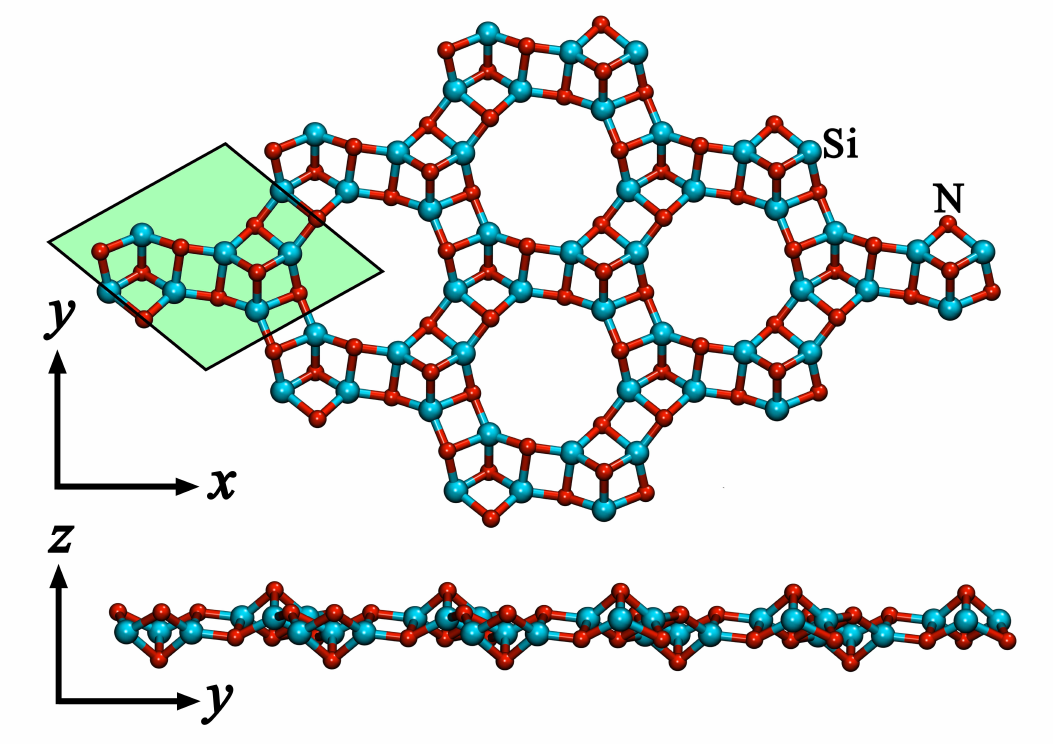}
	\caption{\label{fig:1}
		The $x-y$ and $y-z$ projections of the SLSiN atomic structure, rendered in VMD~\cite{35p} using the Tachyon parallel/multiprocessor ray-tracing engine~\cite{36p}. The green parallelogram highlights the hexagonal Bravais lattice's unit cell, which comprises 6 silicon and 8 nitrogen atoms.}
\end{figure}

Over the past fifty years, the electronic, optical, mechanical, and thermal properties of silicon nitrides family have been comprehensively characterized. To date, nine distinct allotropes of Si$_3$N$_4$--crystallizing in the space groups $P6_3/m$, $P31c$, $I\bar43d$, $Fd\bar3m$, $Pnma$, and $P6_3/mmc$--have been identified. However, a theoretical 2020 study~\cite{me00} predicts a novel single-layer Si$_3$N$_4$ allotrope, designated SLSiN, which exists exclusively in two dimensions (Fig.~\ref{fig:1}).
The discovery arose from our attempt to extract a nanosheet from the hexagonal (3D) $\beta$-Si$_3$N$_4$ parent unit cell, a pursuit inspired by the surge of interest in 2D materials following the landmark isolation of graphene~\cite{graf}.

In this work, we employ density-functional theory (DFT)~\cite{299} to probe the technologically significant yet unexplored influence of charge density modulation on the electronic structure of SLSiN. At its core, this modulation parallels the fundamental process of doping host solids--a longstanding strategy for tuning material properties.

Experimentally, charged two-dimensional (2D) systems have been realizable. A prime example is MoS$_2$ monolayer~\cite{meeee}, where electrons and holes are injected via direct inter-band absorption with a 390 nm pump pulse to investigate charge-carrier dynamics \cite{309}. Understanding how interfacial electronegativity differences induce net monolayer charges is vital for nanotechnology applications--such as field-effect transistors and heterojunction solar cells--where device performance hinges on precisely controlled charge states.

The present work is organized as follows: Sec. II details the computational methodology and simulation setup; Sec. III presents and analyzes the results; and Sec. IV concludes with a summary of our findings.
\subsection*{\label{sec:2}{\normalsize{II. Computational details}}}
We employed a self-consistent~\cite{411} plane-wave pseudopotential framework~\cite{421} within the Perdew--Burke--Ernzerhof (PBE)~\cite{431} formulation of DFT, as implemented in the \qe\ package~\cite{451}. The net charges of the unit cell changes from zero (the neutral or reference system) to $n\,=\,\pm\,3$ (in terms of the elementary electric charge $e$), leading accordingly to seven different systems. The following setup and parameters were applied in this work: Debian-style Linux~\cite{R30,R31} operating system; Open MPI v.3.1.6 for CPU parallelization~\cite{R32}; scalar-relativistic ultrasoft pseudopotentials~\cite{461,471} ({\texttt{N.pbe-n-rrkjus\_psl.1.0.0.UPF}} and {\texttt{Si.pbe-n-rrkjus\_psl.1.0.0.UPF}}~\cite{481}) generated by Rappe--Rabe--Kaxiras--Joannopoulos (RRKJ)~\cite{491} pseudization recipe with nonlinear core correction~\cite{501} to model the core electrons; the valence configurations $2s2p$ and $3s3p$ for N and Si, respectively; the electron density augmented through a Fourier interpolation scheme in real space~\cite{511}; the optimized values of $\sim1225$ and 4900 eV as the kinetic energy cutoff values for the wave functions and for the charge density and the potential, respectively; a vacuum space larger than 15 \AA\ along the $z-$axis to decouple periodic interactions; a conventional approach introducing a neutralizing uniform jellium background charge~\cite{4055} to avoid the divergence of the electrostatic energy arising from the periodic array of charged unit cells; variable-cell relaxations based on the Murnaghan isothermal equation of state~\cite{4155,4255} to find the equilibrium (zero-pressure) lattice constants ($a_0$); fixed-cell relaxations using the Broyden--Fletcher--Goldfarb--Shanno (BFGS)~\cite{521,531} optimization algorithm to find the true atomic positions with a force threshold less than $2.58\,\times\,10^{-4}$ eV.\AA$^{-1}$; Brillouin-zone samplings along the closed $\Gamma$--M--K--$\Gamma$ k-space path meshed by 163 k-points using the MP (Methfessel--Paxton) smearing method~\cite{541}.

All methodological schemes and reported values were chosen according to rigorous energy‐convergence tests; all software tools employed in this study are also either distributed under or compatible with the GNU General Public License (GPL)~\cite{gpl1}.\\
\subsection*{\label{sec:3}{\normalsize{III. Results and discussion}}}
\subsubsection{Equilibrium parameters}
Equilibrium, zero‐pressure lattice constants for all seven monolayers were obtained by fitting their total‐energy versus volume data to the Murnaghan isothermal equation of state, presuming that the bulk modulus $K_0$ responds linearly to pressure under finite strain. We have
\begin{equation}
	\label{eq:pv}
	P(V)\,=\,\frac{K_{0}}{K'_{0}}\left[\left(\frac{V_0}{V}\right)^{K'_{0}}\,-\,1\right],
\end{equation}
where $K'_{0}$ is the first derivative of the bulk modulus with respect to pressure $P$, $V$ is the unit-cell volume, and $V_0$ is the zero-pressure volume. The constancy of $K$ and $K'$ guarantees the linear dependence of the bulk modulus on pressure. Integrating Eq.~\eqref{eq:pv} according to $P(V)\,=\,-dE(V)/dV$ (the first law of thermodynamics at absolute zero) yields
\begin{equation}
	\label{eq:uv}
	E(V)\,=\,E_{gs}\,+\,\frac{{K_{0}}\,V}{K'_{0}}\,\left[\frac{(V_{0}/V)^{K'_{0}}}{{K'_{0}}\,-\,1}\,+\,1\right]\,-\,\frac{{K_{0}}V_0}{{K'_{0}}\,-\,1},
\end{equation} 
where $E$ is the internal energy and $E_{gs}$ is the ground-state energy of the system. After relaxing the atomic positions, a uniform strain is applied to change the lattice constant with an increment of about $\pm\,2.65\times\,10^{-2}$ {\AA}. The total energy of the system is then estimated for each strained structure using self-consistent-field (SCF) calculations. 

The isothermal $P-V$ (Eq.~\ref{eq:pv}) and $E-V$ (Eq.~\ref{eq:uv}) diagrams could then be obtained for each phase, as illustrated in Figs.~\ref{fig:2} and~\ref{fig:3}.
\begin{widetext}
	\begin{minipage}{\linewidth} 
		\begin{figure}[H]
			\centering
			\includegraphics[scale=0.4]{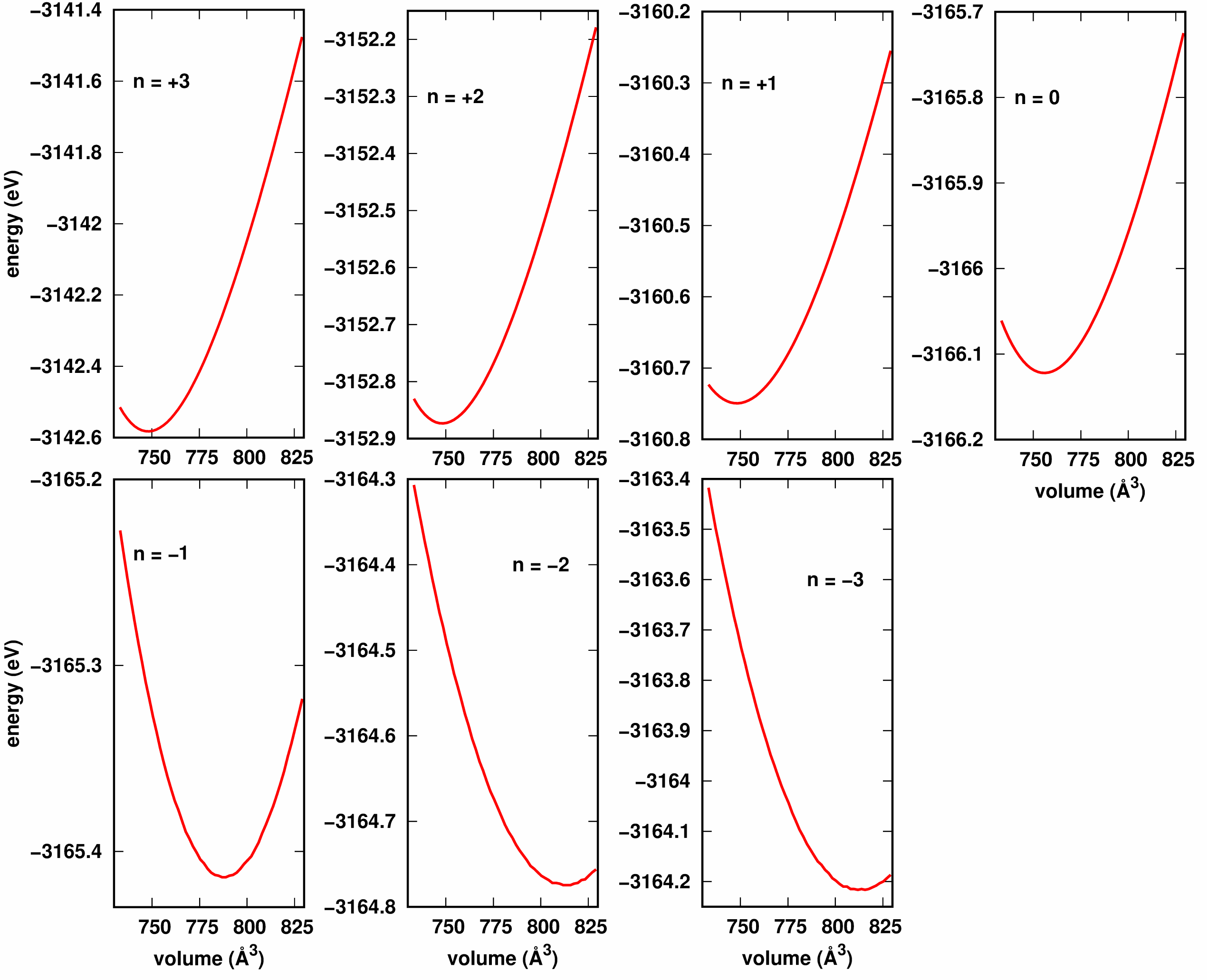}
			\caption{\label{fig:2}
				The internal energy $E$ versus unit-cell volume $V$ for the seven systems at 0 K, computed from Eq.~\ref{eq:uv}--rendered by Gnuplot~\cite{gplt}. The curve minima denote the ground-state energy $E_{gs}$, equilibrium volume $V_0$, and lattice constant $a_0$, as listed in Table~\ref{tab:0}.}
		\end{figure}
	\end{minipage}
\end{widetext}
Each $E-V$ curve exhibits a clear minimum that defines the zero-pressure ground-state energy, equilibrium volume $V_{0}$, and lattice constant $a_{0}$ of the corresponding unit cell.

As the charge state $n$ varies from $+\,3$ to $-\,3$, these minima shift steadily from smaller to larger volumes. The same optimized values can also be obtained from the zero crossings of the $P-V$ curves shown in Fig.~\ref{fig:3}.
\begin{widetext}
	\begin{minipage}{\linewidth} 
		\begin{figure}[H]
			\centering
			\includegraphics[scale=0.35]{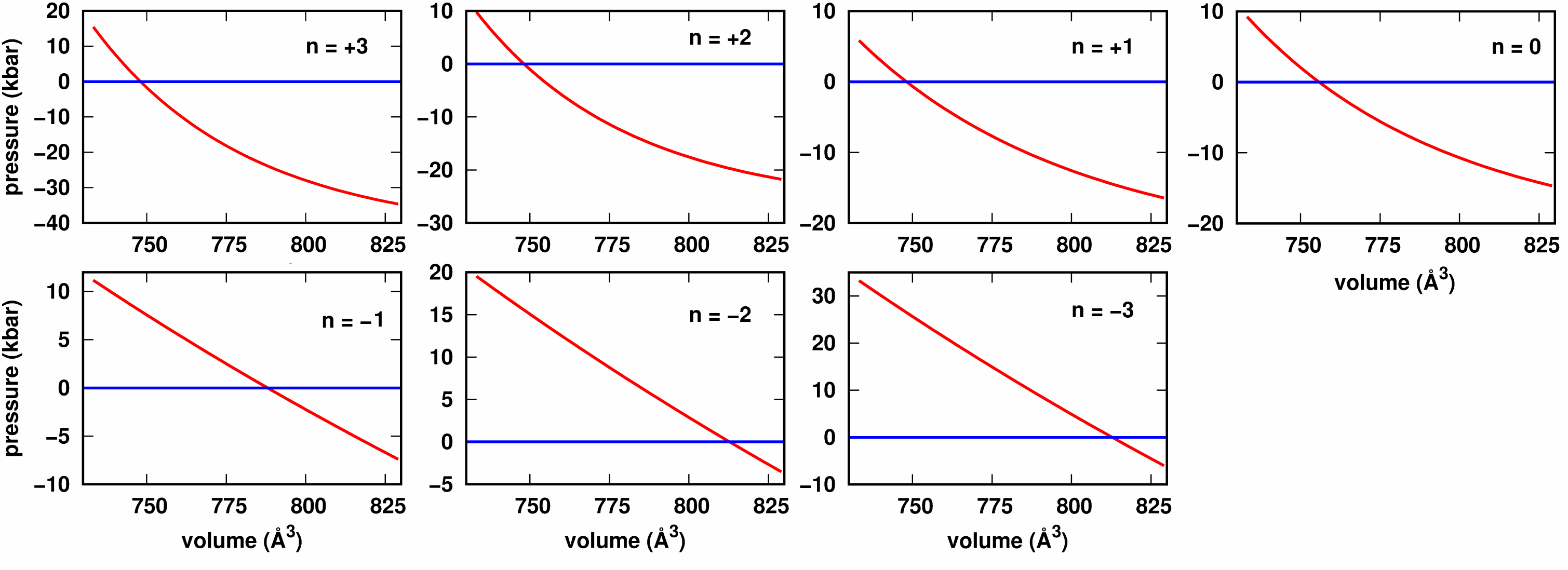}
			\caption{\label{fig:3}
				The pressure $P$ as a function of the unit-cell volume $V$ for the seven SLSiN structures at 0 K, computed via Eq.~\ref{eq:pv}. The points where the pressure curves cross zero mark the ground-state volume $V_{0}$ and lattice constant $a_{0}$, coinciding with the minima of the $E-V$ plots (Fig.~\ref{fig:2}).}
		\end{figure}
	\end{minipage}
\end{widetext}
Such plot-based insights can be rendered fully quantitative by directly examining the underlying numerical data tabulated in Tab.~\ref{tab:0}.
\begin{widetext}
	\begin{minipage}{\linewidth} 
		\begin{table}[H]
			\caption{\label{tab:0}
				The bulk modulus $K_{0}$, its pressure derivative $K'_{0}$, equilibrium volume $V_{0}$, lattice constant $a_{0}$, Si--N bond length, ground-state energy $E_{gs}$, Fermi energy $E_{F}$, and density $\rho$, all obtained by fitting the Murnaghan isothermal equation of state.}
			\begin{tabular}{c|cccccccc}
				\hline
				$\,\,n\,\,$&$\,\,K_0$\,(kbar)\,\,&\,\,\,\,\,\,$K'_0$\,\,\,\,\,\,&\,\,\,\,$V_0$ ({\AA$^3$})\,\,\,\,&\,\,\,\,$a_0$ (\AA)\,\,\,\,&\,\,\,\,Si--N (\AA)\,\,\,\,&\,\,\,\,\,$E_{gs}$ (eV)\,\,\,\,\,&\,\,\,\,\,$E_F$ (eV)\,\,\,\,\,&\,\,$\rho$ (gr/cm$^3$)\,\,\\
				\hline		
				$+\,3$&660.933&15.0&748.0511&7.5589&1.7036&-3142.5828&-11.3273&0.6228\\
				$+\,2$&415.339&15.0&748.1454&7.5591&1.7042&-3152.8728&-9.0729&0.6227\\
				$+\,1$&260.144&10.377&748.1297&7.5591&1.7133&-3160.7491&-6.6307&0.6227\\
				$0$&254.912&11.221&755.8391&7.5850&1.7381&-3166.1220&-3.5031&0.6164\\
				$-\,1$&149.411&1.0&788.0357&7.6912&1.7461&-3165.4145&0.6802&0.5912\\
				$-\,2$&179.826&1.0&812.8197&7.7710&1.7600&-3164.7737&0.5302&0.5732\\
				$-\,3$&306.255&1.0&812.8197&7.7710&1.7615&-3164.2186&0.4724&0.5732\\
				\hline
			\end{tabular}
		\end{table}
	\end{minipage}
\end{widetext}
Both the lattice constant $a_0$ and the equilibrium volume $V_0$ increase monotonically as the charge state shifts from $n\,=\,+\,3$ to $-\,3$. This behavior stems from the extra electrons amplifying Coulomb repulsion between atoms, which stretches the Si--N bonds and, in turn, enlarges the lattice constant. As the unit-cell volume increases, the density $\rho$ also correspondingly decreases according to the fact that the total atomic mass of the unit cell (the combined mass of all the protons and neutrons) is a constant ($\sim\,4.66\,\times\,10^{-25}$ kg). The table also shows that the reference system, $n\,=\,0$, is the most stable one due to having the lowest ground-state energy. With increase in the number of electrons or holes, the system becomes less stable as its energy increases. Consequently, in device applications, SLSiN neither exhibits electronegativity nor electropositivity when interfacing with other materials.

As the charge state $n$ varies from $+\,3$ to $-\,3$, the magnitude of the Fermi energy $E_F$ steadily decreases, which matches the Fermi level of a non-interacting, non-relativistic ensemble (gas) of identical spin-{\scriptsize{1/2}} fermions in three dimensions described by
\begin{equation}
	\label{eq:Fermi}
	E_{F}\,=\,2^{-1}\,3^{2/3}\,\pi^{4/3}\,\left(\frac{N_{0}}{V_0}\right)^{2/3}, \cite{ash}
\end{equation}
in atomic units $(\hbar\,=\,e\,=\,m_e\,=\,1)$, where $N_0$ is the number of electrons in allowed states. As the charge state $(n)$ shifts from $+\,3$ to $-\,3$, both the electron count $N_0$ and equilibrium volume $V_0$ increase; however, the volumetric expansion outstrips the rise in $N_0$, driving an overall reduction in the system's Fermi energy.
\subsubsection{Electronic structure}
Figure \ref{fig:4} presents the electronic band structures of the seven systems at charge states $n\,=\,0,\pm\,1, \pm\,2, \pm\,3$.
\begin{widetext}
	\begin{minipage}{\linewidth} 
		\begin{figure}[H]
			\centering
			\includegraphics[scale=0.5]{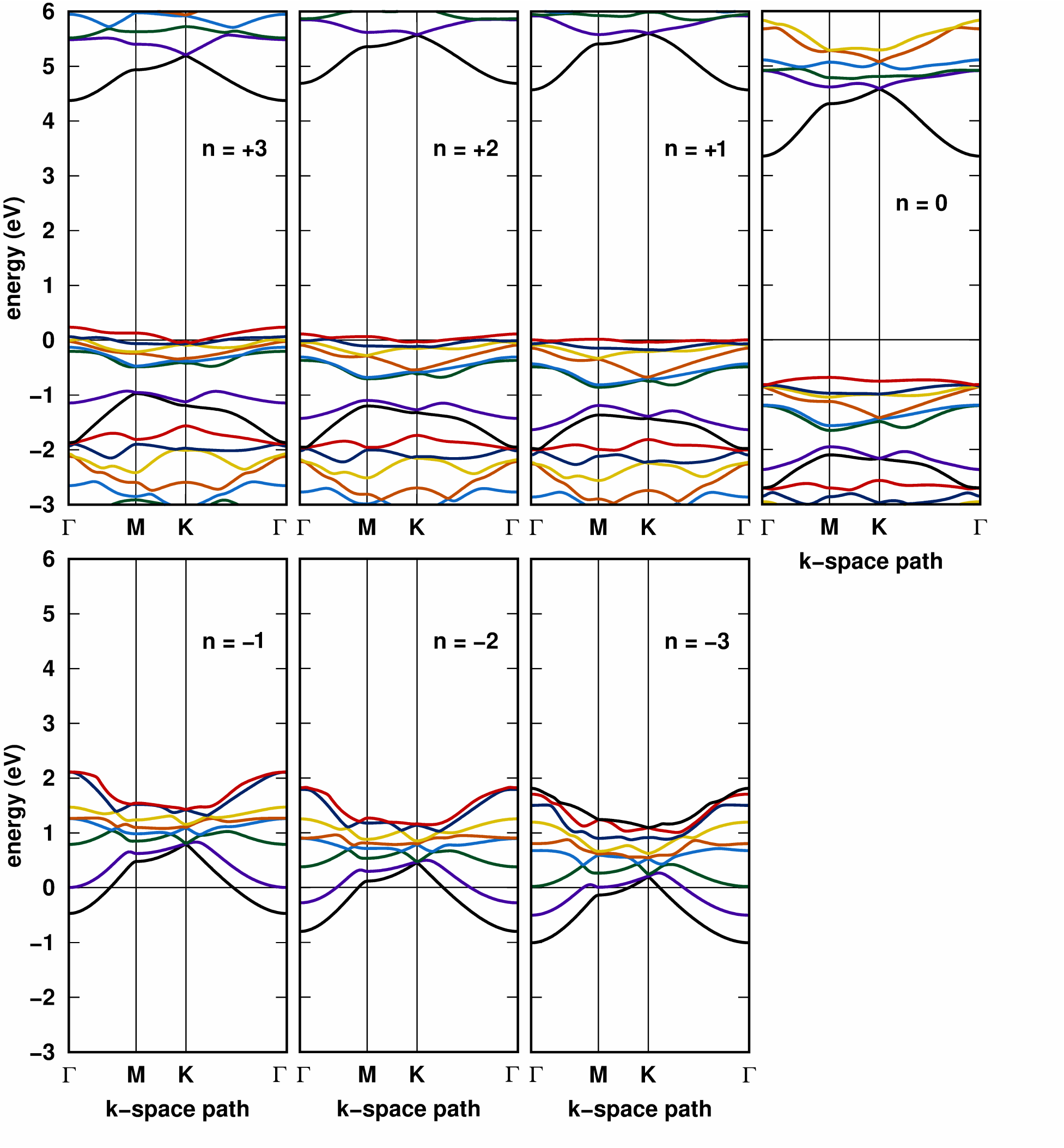}
			\caption{\label{fig:4}
				Electronic band structures of SLSiN for charge states $n\,=\,+\,3$ through $-\,3$. In the neutral cell ($n\,=\,0$), the band gap is about 4 eV; this gap shifts upward as the cell is hole-doped ($n>0$) and downward under electron-doping ($n<0$). All charged configurations ($n\neq0$) exhibit bands crossing the Fermi level, indicating metallic conduction.}
		\end{figure}
	\end{minipage}
\end{widetext}
As is seen, in the neutral state ($n\,=\,0$), the band structure displays an approximately 4 eV gap, confirming its insulating character. However, when the cell is hole-doped ($n\,=\,+\,1$ through $+\,3$) or electron-doped ($n\,=\,-\,1$ through $-\,3$), this gap shifts upward or downward, respectively, causing the bands to cross the Fermi level and render the system metallic. Consequently, every charged configuration behaves as a metal. The narrowing, shifting, and eventual closure of the band gap are likewise evident in the density of states plots shown in Fig.~\ref{fig:5}.
\begin{widetext}
	\begin{minipage}{\linewidth} 
		\begin{figure}[H]
			\centering
			\includegraphics[scale=0.5]{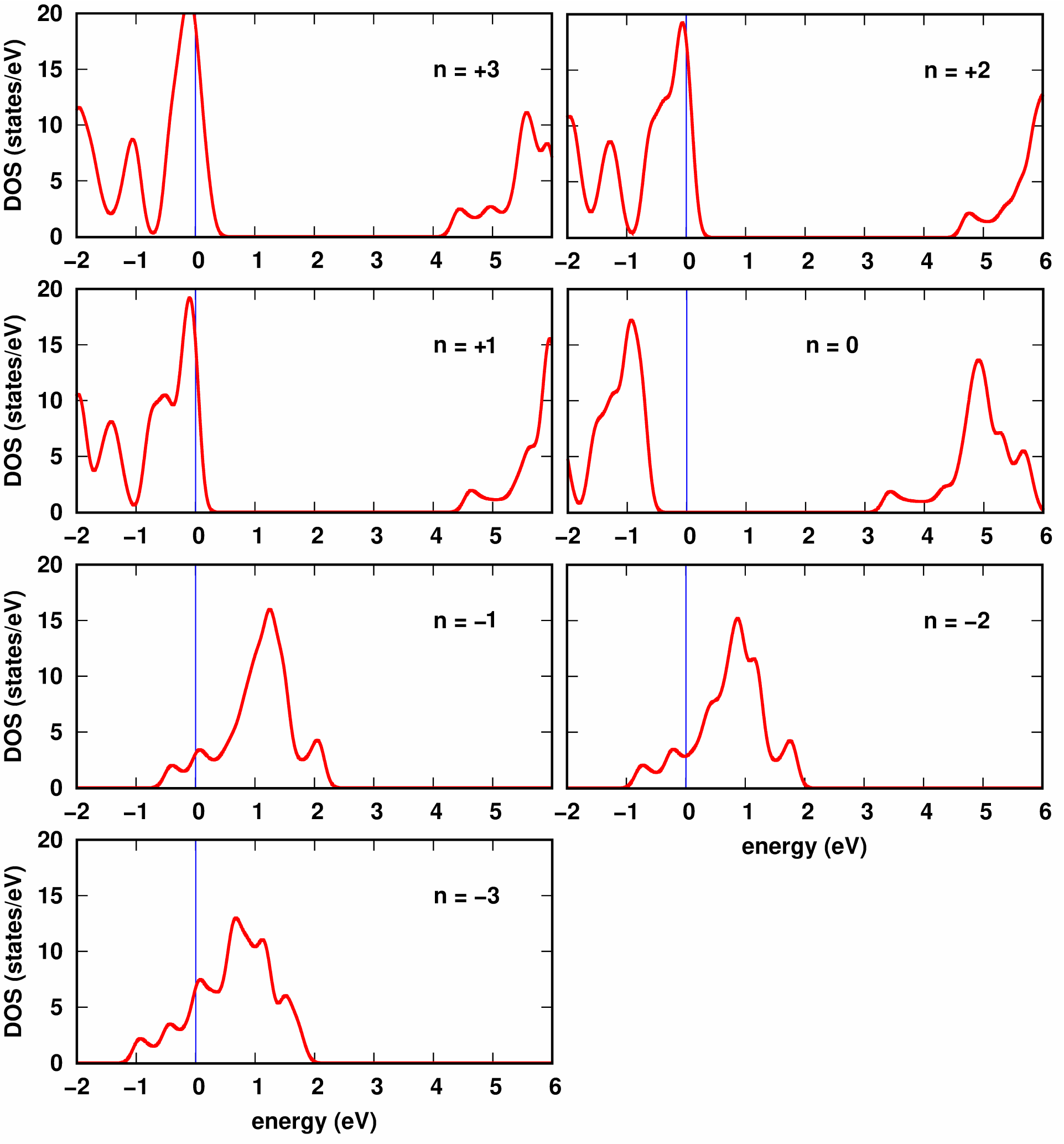}
			\caption{\label{fig:5}
				Electronic density of states (DOS) of SLSiN for unit-cell charges from $+\,3$ to $-\,3$. In the neutral case ($n\,=\,0$), a band gap of about 4 eV appears, which shifts to higher energies under hole doping, and to lower energies under electron doping. All charged configurations exhibit DOS curves crossing the Fermi level, indicating metallic behavior. Moreover, hole-doped cells show a larger DOS at the Fermi level, reflecting superior conductivity compared to electron-doped systems.}
		\end{figure}
	\end{minipage}
\end{widetext}
In the hole-doped cases ($n=+\,1,+\,2,+\,3$), pronounced flat bands appear at the Fermi level, giving rise to sharp DOS peaks and yielding superior electrical conductivity compared with the electron-doped structures. Such flat bands, according to 
\begin{equation}
	\label{eq:emass}
	m^*\,=\,\left(\frac{\partial^2E}{\partial\,k_i\,\partial\,k_j}\right)^{-1},
\end{equation}
indicate high effective mass ($m^*$) or equivalently low group-velocity ($v_g$) regions of the crystal ($v_g\propto\,1/m^*$), in which the charge carriers are more localized, amplifying electronic responses.

Excluding the insulating $n\,=\,0$ case, the DOS at the Fermi level rises monotonically as $n$ goes from $-\,3$ to $+\,3$, indicating that hole doping (electron removal) enhances the monolayer's conductivity far more than electron doping. This behavior stems from the neutral system ($n\,=\,0$) acting as a perfect insulator, with all electronic states completely filled. Thus, hole doping more effectively induces metallicity by creating vacant states that facilitate carrier movement and electrical conduction, whereas adding electrons tends to fill bulk states and limit mobility.
\subsection*{\label{sec:4}{\normalsize{IV. Conclusions}}}
Despite being one of the oldest and most extensively studied material families, silicon nitrides have recently given rise to a novel single-layer Si$_3$N$_4$ allotrope--designated SLSiN--that exists exclusively in 2D form. Here, density-functional theory was used to explore how altering the charge density at the unit-cell level influences the electronic properties of SLSiN. The net charge per cell was varied from $n\,=\,0$ (the neutral reference) to $n\,=\,\pm\,1,\pm\,2$, and $\pm\,3$. The Murnaghan isothermal equation of state was used to find a number of equilibrium, zero-pressure parameters including lattice constant, unit-cell volume, and energy of the seven structures. For each monolayer, electronic band structures and density of states (DOS) were calculated at the PBE level of DFT. In its neutral form ($n\,=\,0$), SLSiN exhibits zero electronegativity, meaning it neither pulls in extra electron density nor sacrifices stability when brought into contact with other materials in a device. In hole-doped SLSiN ($n=+\,1,+\,2,+\,3$), pronounced flat bands appear near the Fermi level, indicating low-group-velocity regions that amplify the material's electronic response. Charging the unit cell from zero to $+/-3$ shifts the original ($n\,=\,0$) bandgap upward or downward, causing the electronic bands to cross the Fermi level; therefore, any configuration with a net charge shows metallic behavior. Excluding the insulating $n\,=\,0$ case, the density of states at the Fermi level increases steadily as nn goes from $n\,=\,-\,3$ to $+\,3$, indicating that hole doping (electron removal) boosts SLSiN's conductivity far more effectively than electron doping.\\
\subsection*{\normalsize{Data availability statement}}
All the data supporting the findings of this study are available within the present article. The associated raw data are also available upon request.

\end{document}